\documentclass[article]{jss}

\usepackage{thumbpdf,lmodern}

\usepackage{framed}

\usepackage{graphicx}
\usepackage{listings}
\usepackage{enumerate} 
\usepackage[section]{placeins}
\usepackage{amsmath}

\usepackage{booktabs}
\usepackage{multirow}
\usepackage{float}

\usepackage{booktabs}
\usepackage[caption = false]{subfig}

\usepackage{hyperref}

\author{Sayan Putatunda\\VMware 
   \And Kiran Rama \\VMware \And Dayananda Ubrangala\\VMware \And  Ravi Kondapalli\\VMware}
\Plainauthor{Sayan Putatunda, Kiran Rama, Dayananda Ubrangala, Ravi Kondapalli}

\title{SmartEDA: An \proglang{R} Package for Automated Exploratory Data Analysis}
\Plaintitle{SmartEDA: An R Package for Automated Exploratory Data Analysis}
\Shorttitle{SmartEDA: An R Package for Automated Exploratory Data Analysis}

\Abstract{
 This paper introduces SmartEDA, which is an \proglang{R} package for performing Exploratory data analysis (EDA). EDA is generally the first step that one needs to perform before developing any machine learning or statistical models. The goal of EDA is to help someone perform the initial investigation to know more about the data via descriptive statistics and visualizations. In other words, the objective of EDA is to summarize and explore the data. The need for EDA became one of the factors that led to the development of various statistical computing packages over the years including the R programming language that is a very popular and currently the most widely used software for statistical computing. However, EDA is a very tedious task, requires some manual effort  and some of the open source packages available  in R are not just upto the mark. In this paper, we propose a new open source package i.e. SmartEDA for R to address the need for automation of exploratory data analysis. We discuss the various features of SmartEDA and illustrate some of its applications for generating actionable insights using a couple of real-world datasets. We also perform a comparative study of SmartEDA with respect to other packages available for exploratory data analysis in the Comprehensive R Archive Network (CRAN). 
}

\Keywords{Exploratory Data Analysis, Data Mining, Machine Learning, Statistical Learning, Descriptive Analytics, \proglang{R}}
\Plainkeywords{Exploratory Data Analysis, Data Mining, Machine Learning, Statistical Learning, Descriptive Analytics, R}

\Address{
  Sayan Putatunda, Kiran Rama, Dayananda Ubrangala, Ravi Kondapalli\\
  VMware Software India Pvt Ltd.\\
  Kalyani Vista, Phase 4, JP Nagar, 
  Bengaluru, Karnataka 560076\\
  Phone: 080 4044 0000 \\
  E-mail: \email{sayanp@iima.ac.in, efpm04013@iiml.ac.in, \\daya6489@gmail.com, rpkondapalli@yahoo.com }\\
  URL: \url{https://www.vmware.com/in.html}
}

\begin{document}



\section[Introduction: Exploratory Data Analysis]{Introduction: Exploratory Data Analysis} \label{intro}
Nowadays, we see applications of Data Science almost everywhere. Some of the well highlighted aspects of data science are the various statistical and machine learning techniques applied for solving a problem. However, any data science activity starts with an Exploratory Data Analysis (EDA). The term "Exploratory Data Analysis" was coined by \cite{eda:1}. EDA can be defined as the art and science of performing initial investigation on the data by means of statistical and visualization techniques that can bring out the important aspects in the data that can be used for further analysis \citep{eda:1}. EDA puts an emphasis on hypothesis generation and pattern recognition from the raw data \citep{eda:8}. There have been many studies conducted on EDA reported in the Statistics literature (please see Section \ref{lit} for more details).

\begin{figure}[!htp]
\centering
\includegraphics[width=0.75\textwidth]{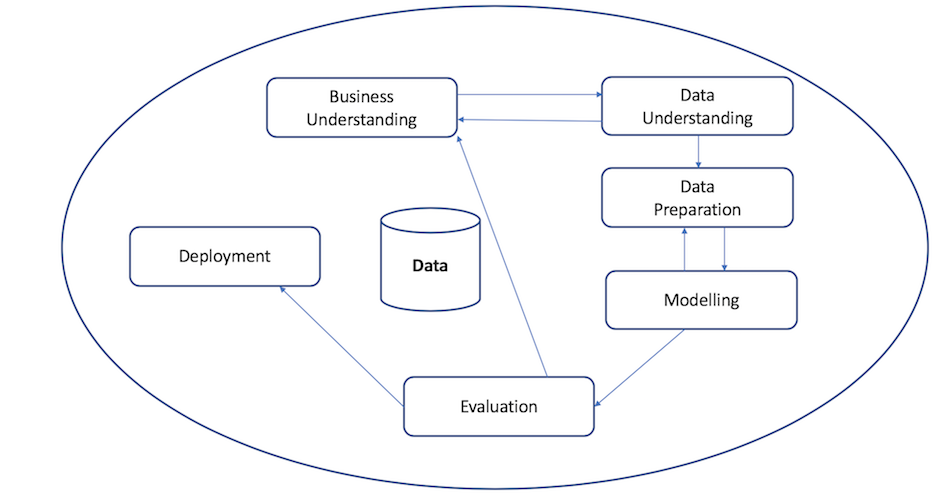}
\caption{The CRISP-DM Process Model for Data Mining \citep{crisp:1}}
\label{fig:crisp}
\end{figure}

EDA is a very important component of the Data Mining process as per the industry standard CRSP-DM framework. The CRISP-DM stands for "CRoss Industry Standard Process for Data Mining" \citep{crisp:1}. Data mining is a creative process that required a different set of skills and knowledge. However, earlier there was a lack of any standard framework for Data mining projects, which esured that the sucess/failure of a data mining project is highly dependent on the skill-set of a particular individual or a team that is executing the project. To address this need, \cite{crisp:1} proposed the CRISP-DM process mopdel that  is a framework for executing any data mining project. The CRISP-DM framework is independent of the tools used and the industry sector. 

Figure \ref{fig:crisp} shows the different components (viz. Business understanding, Data understanding, Data preparation, Modeling, Evaluation and Deployment) of the CRISP-DM process model for data mining. It is a cyclic process where there is a feedback loop between some components. We can see that the "Data Understanding" is a very important component which affects the "Business Understanding" as well. EDA helps in Data understanding and thus directly impacts the quality and success of a data mining project.

EDA can be categorized into Descriptive statistical techniques and graphical techniques \citep{eda:11}. The first category encompasses various univariate and multivariate statistical techniques whereas the second category comprises the various visualization techniques. Both of these techniques are used to explore the data, understand the patterns in the data, understand the existing relationships between the variables and most importantly, generate data drive insights that can be used by the business stakeholders. However, EDA requires a lot of manual effort and also a substantial amount of coding effort in statistical computing packages such as R \citep{Venables+Ripley:2002}. There is a huge need for automation of the EDA process and this motivated us to develop the SmartEDA package and come up with this paper. 

The contribution of this paper is in development of a novel R package i.e. SmartEDA that addresses the need for automating the EDA process. The main benefits of SmartEDA are in development time savings, less error percentage and reproducibility. Although there are other packages available in the Comprehensive R Archive Network (CRAN)  for EDA (such as Hmisc, DataExplorer and more) but SmartEDA has additional functionalities when compared to them including extension to data.table package, capability for performing summary statistics and ability to plot for both numerical and categorical variables and many more (please see Section \ref{compare} for more details).

The rest of this paper is structured as follows. In Section \ref{lit}, we give a brief review of the literature. Section \ref{smarteda} gives an overview of the SmartEDA package available in CRAN and its various functionalities. In Section \ref{illustrate}, we apply SmartEDA to generate actionable insights for a couple of real world datasets. We then follow it up with Section \ref{compare} where we compare SmartEDA with some of the other packages for EDA available in CRAN. Finally, Section \ref{con} concludes this paper.

\section{Related Work} \label{lit}
Some of earliest work done on Exploratory Data Analysis (EDA) including coining the term and defining some of the basic EDA techniques was done by \cite{eda:1}. However, many researchers have formulated different definitions of EDA over the years. One of the widely accepted definition is that "Exploratory data analysis isolates patterns and features of the data and reveals these forcefully to the analyst" \citep{eda:12}. \cite{eda:2} formulated the four basic elements of EDA namely, (a) residual, (b) re-expression, (c) resistant and (d) display. \cite{eda:3} worked on this framework and updated the four elements with relevant techniques and they re-named "display" as "revelation". \cite{eda:6} contrasted Exploratory Data Analysis (EDA) with Confirmatory Data Analysis (CDA) and proposed that EDA complements CDA. \cite{eda:5} proposed an unified approach to confirmatory data analysis and exploratory data analysis using graphical data displays.

\cite{eda:4} introduced EDA in the context of data mining and resampling with focus on pattern recognition, cluster detection and variable selection. Over the years, EDA has been used various applications across different domains such as geoscience research \citep{eda:7}, auditing \citep{eda:8}, game-based assessments \citep{eda:9}, clinical study groups \citep{eda:10} and more.

Moreover, there have been various packages developed in R for performing EDA on raw data such as dlookr \citep{pack:1}, DataExplorer \citep{pack:2}, Hmisc \citep{pack:3}, exploreR \citep{pack:4}, RtutoR \citep{pack:5} and summarytools \citep{pack:6}.

\section{An overview of the SmartEDA package} \label{smarteda}
The SmartEDA R package is publicly available in the  Comprehensive R Archive Network (CRAN) \citep{smartEDA}.  It has got more than $3100+$ downloads as of March 2019, which indicates its acceptability and maturity in the Statistics and Machine learning community.

The SmartEDA package automatically select the variables and performs the related descriptive statistics. Moreover, it also analyzes information value, weight of evidence, custom tables, summary statistics and performs graphical techniques for both numeric and categorical variables. 

Some of the most important advantages of the SmartEDA package are that it can help in applying end to end EDA process without having to remember the different R package names, write lengthy R scripts, no manual effort required to prepare the EDA report and finally, automatically categorize the variables  into the right data type (viz. Character, Numeric, Factor and more) based on the input data. Thus, the main benefits of SmartEDA are in development time savings, less error percentage and reproducibility. 

Moreover, the SmartEDA package has customized options for the data.table package such as (1) Generate appropriate summary statistics depending on the data type, (2) Data reshaping using data.table.dcast(), (3) Filter rows/cases where conditions are true. Options to apply filters at variable level or complete data set like base subsetting and (4) Options to calculate measures of central tendency (like Mean, Median, Mode, etc.), measures of variance/dispersion (like Standard Deviation, Variance, etc.), Count, Proportions, Quantiles, IQR, Percentages of Shares (PS) for numerical data.

\begin{figure}[!htp]
\centering
\includegraphics[width=0.9\textwidth]{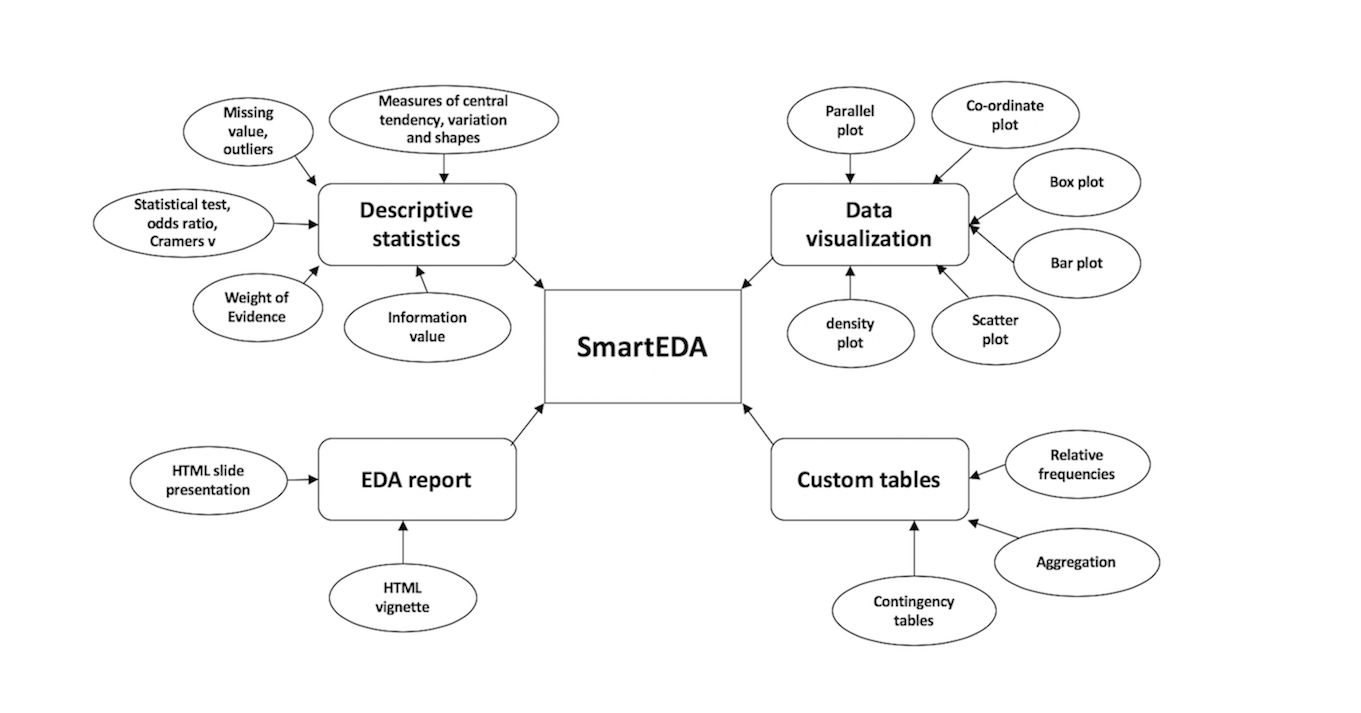}
\caption{The various functionalities of SmartEDA}
\label{fig:features}
\end{figure}

\begin{table}[!htp]
\centering
\caption{Functionalities and the corresponding R functionss of SmartEDA}
\label{tab:concise3}
\scalebox{0.75}{
\begin{tabular}{ll}
\toprule
Functionalities        & SmartEDA functions              \\ \midrule
Descriptive statistics & ExpData(), ExpNumStat(),  ExpCatStat()        \\ \noalign{\smallskip}\hline
Data visualization     & ExpCatViz(), ExpNumViz(), ExpParcoord(), ExpOutQQ() \\ \noalign{\smallskip}\hline
Custom table           & ExpCustomStat()            \\ \noalign{\smallskip}\hline
HTML EDA report        & ExpReport()  \\ \bottomrule
\end{tabular}}
\end{table}

Figure \ref{fig:features} summarizes the various functionalities of SmartEDA. The SmartEDA R package has four unique functionalities as described in Table \ref{tab:concise3}. To know more about the specific commands/functions to execute the above mentioned functions of SmartEDA, please refer to the detailed package documentation available in CRAN \citep{smartEDA}.

\section{Illustrations} \label{illustrate}
In this section, we illustrate the various functionalities of SmartEDA to generate actionable insights for couple of publicly available datasets namely, Carseats and NYC flights data. It is to be noted that we have used the version $3$ of the SmartEDA package for all the illustrations in this paper.

\subsection{EDA for Sales of Child Car seats at different locations}
We apply SmartEDA to generate insights on the sales of Child car seats at different locations. We will use the "Carseats" data available in the ISLR package \citep{pack:7} that contains $11$ variables such as unit sales in each locations (Sales), price charged by competitors (CompPrice), community income level, (Income) population size in region (population), advertising budget (Advertising), price company charges for car seats in each site (Price), quality of shelving location (ShelveLoc), average age of local population (Age), education level at each location (Education), urban/rural location indicator (Urban) and US store/non-US store indicator (US).

We will now use SmartEDA for understanding the dimensions of the dataset, variable names, overall missing summary and data types of each variables.
\begin{CodeChunk}
\begin{CodeInput}
R> library("SmartEDA")
R> library("ISLR")
R> Carseats <- ISLR::Carseats
\end{CodeInput}
\end{CodeChunk}
\begin{CodeChunk}
\begin{CodeInput}
R> ExpData(data=Carseats,type=1)
\end{CodeInput}
\begin{CodeOutput}
                               Descriptions        Obs
1                         Sample size (Nrow)       400
2                    No. of Variables (Ncol)        11
3                   No. of Numeric Variables         8
4                    No. of Factor Variables         3
5                      No. of Text Variables         0
6                   No. of Logical Variables         0
7                      No. of Date Variables         0
8   No. of Zero variance Variables (Uniform)         0
9      %. of Variables having complete cases        100% (11)
10 %. of Variables having <50% missing cases         0% (0)
11 %. of Variables having >50% missing cases         0% (0)
12 %. of Variables having >90% missing cases         0% (0)
\end{CodeOutput}
\end{CodeChunk}

\begin{CodeChunk}
\begin{CodeInput}
R> ExpNumStat(Carseats,by="A",gp=NULL,Qnt=NULL,MesofShape=2,
Outlier=FALSE,round=2,Nlim=10)
\end{CodeInput}
\end{CodeChunk}

Now let us look at the summary of the numerical/integer variables such as Advertising, Age, CompPrice, Income, Population, Price and Sales.

\begin{table}[!htp]
\centering
\caption{Summary of numerical variables of Carseats data}
\label{tab:sum1}
\scalebox{0.48}{
\begin{tabular}{llllllllllllllllllll}
\toprule
Vname       & Group & TN  & nNeg & nZero & nPos & NegInf & PosInf & NA Value & Per of Missing & sum     & min & max   & mean   & median & SD     & CV   & IQR   & Skweness & Kurtosis \\ \midrule
Advertising & All   & 400 & 0    & 144   & 256  & 0      & 0      & 0        & 0              & 2654    & 0   & 29    & 6.63   & 5      & 6.65   & 1    & 12    & 0.64     & -0.55    \\
Age         & All   & 400 & 0    & 0     & 400  & 0      & 0      & 0        & 0              & 21329   & 25  & 80    & 53.32  & 54.5   & 16.2   & 0.3  & 26.25 & -0.08    & -1.14    \\
CompPrice   & All   & 400 & 0    & 0     & 400  & 0      & 0      & 0        & 0              & 49990   & 77  & 175   & 124.97 & 125    & 15.33  & 0.12 & 20    & -0.04    & 0.03     \\
Income      & All   & 400 & 0    & 0     & 400  & 0      & 0      & 0        & 0              & 27463   & 21  & 120   & 68.66  & 69     & 27.99  & 0.41 & 48.25 & 0.05     & -1.09    \\
Population  & All   & 400 & 0    & 0     & 400  & 0      & 0      & 0        & 0              & 105936  & 10  & 509   & 264.84 & 272    & 147.38 & 0.56 & 259.5 & -0.05    & -1.2     \\
Price       & All   & 400 & 0    & 0     & 400  & 0      & 0      & 0        & 0              & 46318   & 24  & 191   & 115.8  & 117    & 23.68  & 0.2  & 31    & -0.12    & 0.43     \\
Sales       & All   & 400 & 0    & 1     & 399  & 0      & 0      & 0        & 0              & 2998.53 & 0   & 16.27 & 7.5    & 7.49   & 2.82   & 0.38 & 3.93  & 0.18     & -0.1                           \\ \bottomrule
\end{tabular}}
\end{table}
We will now check for the summary of categorical variables namely, ShelveLoc, Urban and US.
\begin{CodeChunk}
\begin{CodeInput}
R> ExpCTable(Carseats)
\end{CodeInput}
\begin{CodeOutput}
    Variable  Valid Frequency Percent CumPercent
1  ShelveLoc    Bad        96   24.00      24.00
2  ShelveLoc   Good        85   21.25      45.25
3  ShelveLoc Medium       219   54.75     100.00
4  ShelveLoc  TOTAL       400      NA         NA
5      Urban     No       118   29.50      29.50
6      Urban    Yes       282   70.50     100.00
7      Urban  TOTAL       400      NA         NA
8         US     No       142   35.50      35.50
9         US    Yes       258   64.50     100.00
10        US  TOTAL       400      NA         NA
\end{CodeOutput}
\end{CodeChunk}

\begin{figure}[!htp]
    \centering
    \subfloat[Scatter plot of Price vs. Population]{{\includegraphics[width=0.5\textwidth, keepaspectratio]{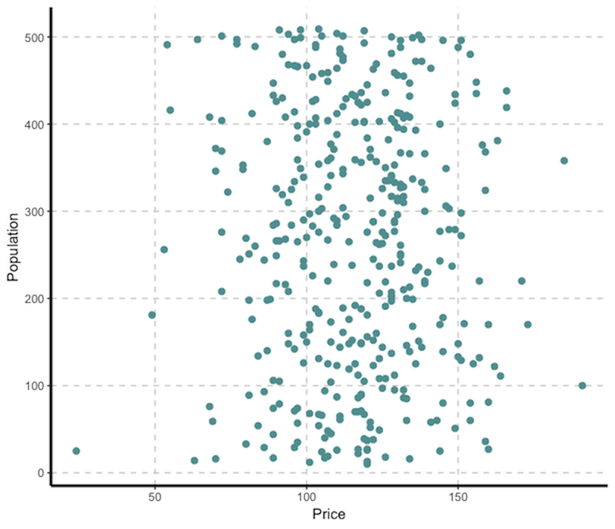} }}%
    \qquad
    \subfloat[Density plot for Income]{{\includegraphics[width=0.5\textwidth, keepaspectratio]{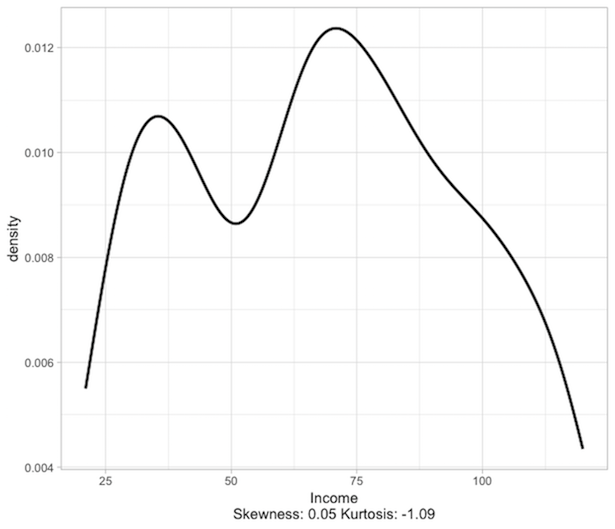} }}%
    \caption{EDA on Carseats data: Scatter plot and Density plot}%
    \label{fig:cars1}%
\end{figure}
\begin{figure}[!htp]
    \centering
    \subfloat[Bar plot for Urban]{{\includegraphics[width=0.5\textwidth, keepaspectratio]{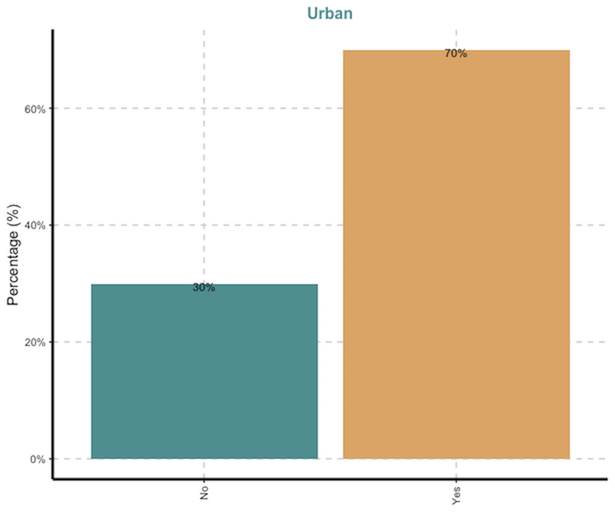} }}%
    \qquad
    \subfloat[Box plot for Population vs. US]{{\includegraphics[width=0.5\textwidth, keepaspectratio]{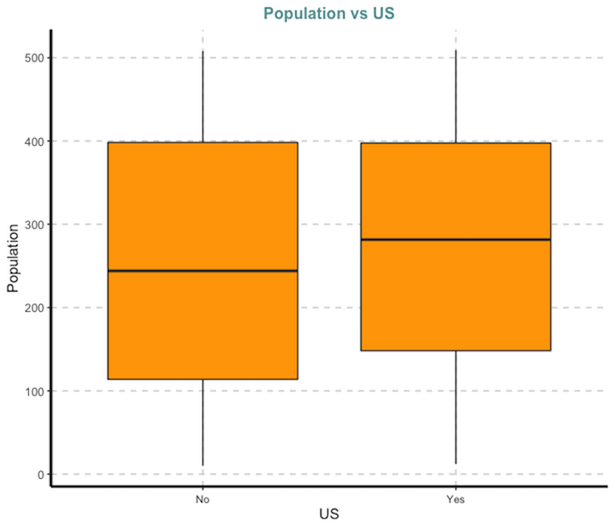} }}%
    \caption{EDA on Carseats data: Bar plot and Box plot}%
    \label{fig:cars2}%
\end{figure}
\begin{figure}[!htp]
    \centering
    \subfloat[Normality plot for Sales]{{\includegraphics[width=0.5\textwidth, keepaspectratio]{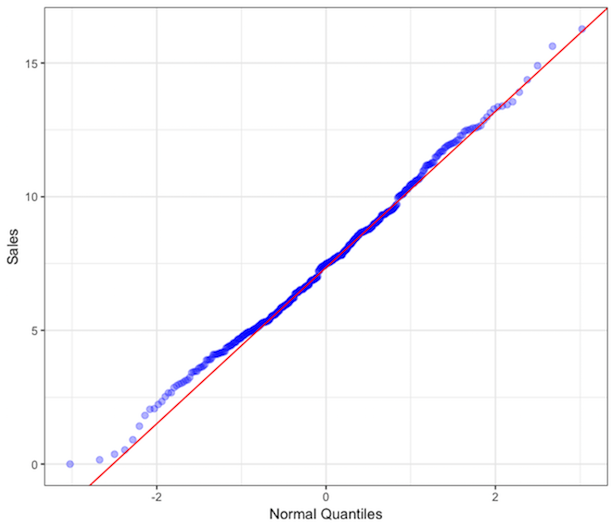} }}%
    \qquad
    \subfloat[Co-ordinate plot for Price, Income, Advertising, Population, Age and Education with respect to the Shelving location]{{\includegraphics[width=0.5\textwidth, keepaspectratio]{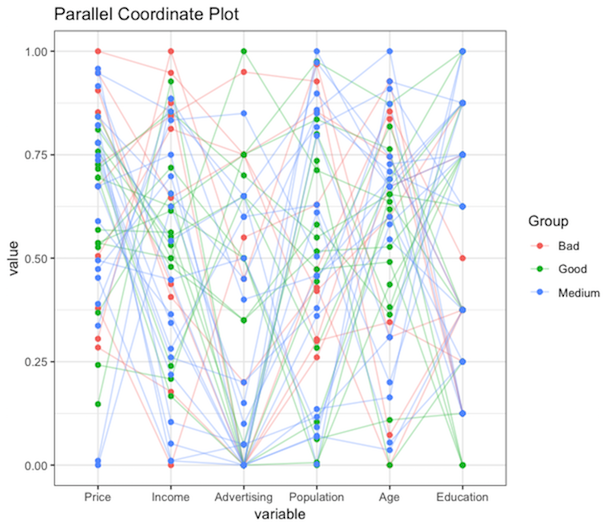} }}%
    \caption{EDA on Carseats data: Normality plot and Co-ordinate plot}%
    \label{fig:cars3}%
\end{figure}

We can visualize the different graphical representations of the data using the SmartEDA package. Figures \ref{fig:cars1}, \ref{fig:cars2} and \ref{fig:cars3} show the different graphical visualizations (such as Scatter plot, Density plot, Bar plot, Box plot, Normality plot and Co-ordinate plot) applied on the "Carseats" dataset.

\begin{CodeChunk}
\begin{CodeInput}
## Scatter plot
R> ExpNumViz(Carseats,gp="Price",nlim=4,fname=NULL,col=NULL,Page=NULL,sample=1) 
## Density plot
R> ExpNumViz(Carseats,gp=NULL,nlim=10,sample=1) 
## Bar plot
R> ExpCatViz(Carseats,gp=NULL,clim=5,margin=2,sample=1) 
## Box plot
R> ExpNumViz(Carseats,gp="US",type=2,nlim=10,sample=1) 
## Normality plot
R> ExpOutQQ(Carseats,nlim=10,sample=1)
## Co-ordinate plots
R>ExpParcoord(Carseats,Group="ShelveLoc",Stsize=c(10,15,20),Nvar=
             c("Price","Income","Advertising","Population","Age","Education")) 
\end{CodeInput}
\end{CodeChunk}

Some of the above functions such as, ExpNumViz, ExpCatViz and ExpOutQQ if executed without the "sample" argument then we will get all the possible plot with various combinations of the relevant variables. For example, ExpNumViz() and ExpCatViz() will generate the required plots for all the possible combinations of numerical variables and categorical variables respectively. Similarly, ExpOutQQ() will generate the normality plot for all the variables in the Carseats dataset. Here, $sample= 1$ argument represents that we need the function to display one plot only. Moreover, the ExpParcoord() function plots the parallel coordinate plot for the mentioned variables such as Price, Income and more. The "Stsize" argument represents the stratified sample size for each class of the group variable i.e. the Shelving location ("ShelveLoc"). The parallel coordinate plot is generally used to detect outliers.

Some of the key insights that we can get from the Figures \ref{fig:cars1}, \ref{fig:cars2} and \ref{fig:cars3} are that there is no correlation between price and population. Most of the related population are from Urban regions and from US. The variable "Sales" is normally distributed. 

\subsection{EDA for NYC Flights Departure Data}
We apply SmartEDA to generate insights on the airline on-time data for all flights departing NYC in 2013. We will use the various datasets namely, flights, airlines, planes and airports available in the nycflights13 package \citep{pack:8} and combine these datasets into a single dataset i.e. "flight\_data" as shown in the following code chunk.

\begin{CodeChunk}
\begin{CodeInput}
R> library("nycflights13")

R> merge_airlines <- merge(flights, airlines, by = "carrier", all.x = TRUE)

R> merge_planes <- merge(merge_airlines, planes, by = "tailnum",
 all.x = TRUE, suffixes = c("_flights", "_planes"))
 
R> merge_airports_origin <- merge(merge_planes, airports, by.x = "origin", 
by.y = "faa", all.x = TRUE, suffixes = c("_carrier", "_origin"))

R> flight_data <- merge(merge_airports_origin, airports, by.x = "dest",
 by.y = "faa", all.x = TRUE, suffixes = c("_origin", "_dest"))
\end{CodeInput}
\end{CodeChunk}

The flight\_data dataset contains around $42$ variables such as flight departure date,  departure/arrival times, carrier information, flight number, tail number, time spent in air, distance between airports, latitude and longitude of airports, flight altitude, timezone offset, airline names, flight speed, type of pane, year of manufacture, number of seats/engine, type of engine and more. Please see the nycflights13 package documentation \citep{pack:8} for a detailed description of all the variables.

We will now use SmartEDA for understanding the dimensions of the dataset, variable names, overall missing summary and data types of each variables.

\begin{CodeChunk}
\begin{CodeInput}
R> library("SmartEDA")
R> ExpData(data=flight_data,type=1)
\end{CodeInput}
\begin{CodeOutput}
                                Descriptions         Obs
1                         Sample size (Nrow)      336776
2                    No. of Variables (Ncol)          42
3                   No. of Numeric Variables          26
4                    No. of Factor Variables           0
5                      No. of Text Variables          15
6                   No. of Logical Variables           0
7                      No. of Date Variables           2
8   No. of Zero variance Variables (Uniform)           4
9      %. of Variables having complete cases    50% (21)
10 %. of Variables having <50% missing cases 47.62% (20)
11 %. of Variables having >50% missing cases      0% (0)
12 %. of Variables having >90% missing cases   2.38% (1)
\end{CodeOutput}
\end{CodeChunk}

Now let us look at the summary of the numerical/integer variables such as time in air, altitude, delay in arrival, arrival time, departure time, delay in departure, flight details, location coordinates (i.e. latitude and longitude), speed, number of seats and more.

\begin{CodeChunk}
\begin{CodeInput}
R> ExpNumStat(flight_data,by="A",gp=NULL,Qnt=NULL,
      MesofShape=2,Outlier=FALSE,round=2,Nlim=10)
\end{CodeInput}
\end{CodeChunk}

\begin{table}[!htp]
\centering
\caption{Summary of numerical variables of NYC flights data}
\label{tab:sum2}
\scalebox{0.4}{
\begin{tabular}{llllllllllllllllllll}
\toprule
Vname            & Group & TN     & nNeg   & nZero & nPos   & NegInf & PosInf & NA\_Value & Per\_of\_Missing & sum          & min     & max    & mean    & median & SD      & CV    & IQR   & Skweness & Kurtosis \\ \midrule
air\_time        & All   & 336776 & 0      & 0     & 327346 & 0      & 0      & 9430      & 2.8              & 49326610     & 20      & 695    & 150.69  & 129    & 93.69   & 0.62  & 110   & 1.07     & 0.86     \\
alt\_dest        & All   & 336776 & 0      & 0     & 329174 & 0      & 0      & 7602      & 2.26             & 191953920    & 3       & 6602   & 583.14  & 433    & 937.61  & 1.61  & 722   & 3.82     & 16.53    \\
arr\_delay       & All   & 336776 & 188933 & 5409  & 133004 & 0      & 0      & 9430      & 2.8              & 2257174      & -86     & 1272   & 6.9     & -5     & 44.63   & 6.47  & 31    & 3.72     & 29.23    \\
arr\_time        & All   & 336776 & 0      & 0     & 328063 & 0      & 0      & 8713      & 2.59             & 492768669    & 1       & 2400   & 1502.05 & 1535   & 533.26  & 0.36  & 836   & -0.47    & -0.19    \\
day              & All   & 336776 & 0      & 0     & 336776 & 0      & 0      & 0         & 0                & 5291016      & 1       & 31     & 15.71   & 16     & 8.77    & 0.56  & 15    & 0.01     & -1.19    \\
dep\_delay       & All   & 336776 & 183575 & 16514 & 128432 & 0      & 0      & 8255      & 2.45             & 4152200      & -43     & 1301   & 12.64   & -2     & 40.21   & 3.18  & 16    & 4.8      & 43.95    \\
dep\_time        & All   & 336776 & 0      & 0     & 328521 & 0      & 0      & 8255      & 2.45             & 443210949    & 1       & 2400   & 1349.11 & 1401   & 488.28  & 0.36  & 837   & -0.02    & -1.09    \\
distance         & All   & 336776 & 0      & 0     & 336776 & 0      & 0      & 0         & 0                & 350217607    & 17      & 4983   & 1039.91 & 872    & 733.23  & 0.71  & 887   & 1.13     & 1.19     \\
flight           & All   & 336776 & 0      & 0     & 336776 & 0      & 0      & 0         & 0                & 664096549    & 1       & 8500   & 1971.92 & 1496   & 1632.47 & 0.83  & 2912  & 0.66     & -0.85    \\
hour             & All   & 336776 & 0      & 0     & 336776 & 0      & 0      & 0         & 0                & 4438791      & 1       & 23     & 13.18   & 13     & 4.66    & 0.35  & 8     & 0        & -1.21    \\
lat\_dest        & All   & 336776 & 0      & 0     & 329174 & 0      & 0      & 7602      & 2.26             & 11858183.47  & 21.32   & 61.17  & 36.02   & 36.1   & 5.73    & 0.16  & 8.51  & -0.27    & -0.92    \\
lon\_dest        & All   & 336776 & 329174 & 0     & 0      & 0      & 0      & 7602      & 2.26             & -29455697.52 & -157.92 & -68.83 & -89.48  & -83.35 & 14.96   & -0.17 & 15.19 & -1.23    & 0.73     \\
minute           & All   & 336776 & 0      & 60696 & 276080 & 0      & 0      & 0         & 0                & 8833668      & 0       & 59     & 26.23   & 29     & 19.3    & 0.74  & 36    & 0.09     & -1.24    \\
month            & All   & 336776 & 0      & 0     & 336776 & 0      & 0      & 0         & 0                & 2205381      & 1       & 12     & 6.55    & 7      & 3.41    & 0.52  & 6     & -0.01    & -1.19    \\
sched\_arr\_time & All   & 336776 & 0      & 0     & 336776 & 0      & 0      & 0         & 0                & 517415985    & 1       & 2359   & 1536.38 & 1556   & 497.46  & 0.32  & 821   & -0.35    & -0.38    \\
sched\_dep\_time & All   & 336776 & 0      & 0     & 336776 & 0      & 0      & 0         & 0                & 452712768    & 106     & 2359   & 1344.25 & 1359   & 467.34  & 0.35  & 823   & -0.01    & -1.2     \\
seats            & All   & 336776 & 0      & 0     & 284170 & 0      & 0      & 52606     & 15.62            & 38851317     & 2       & 450    & 136.72  & 149    & 71.8    & 0.53  & 134   & 0.15     & 0.21     \\
speed            & All   & 336776 & 0      & 0     & 963    & 0      & 0      & 335813    & 99.71            & 145206       & 90      & 432    & 150.79  & 126    & 95.14   & 0.63  & 22    & 2.39     & 4.28     \\
year\_planes     & All   & 336776 & 0      & 0     & 278864 & 0      & 0      & 57912     & 17.2             & 558117792    & 1956    & 2013   & 2001.4  & 2002   & 6.39    & 0     & 7     & -0.87    & 1.92                          \\ \bottomrule
\end{tabular}}
\end{table}

The following code chunk shows the functions in SmartEDA that are applied on the NYC flights data to get a summary of all the categorical and character variables.

\begin{CodeChunk}
\begin{CodeInput}
R> ExpCTable(flight_data)
\end{CodeInput}
\begin{CodeOutput}
      Variable                    Valid Frequency Percent CumPercent
1       origin                      EWR    120835   35.88      35.88
2       origin                      JFK    111279   33.04      68.92
3       origin                      LGA    104662   31.08     100.00
4       origin                    TOTAL    336776      NA         NA
5         type  Fixed wing multi engine    282074   83.76      83.76
6         type Fixed wing single engine      1686    0.50      84.26
7         type                       NA     52606   15.62      99.88
8         type               Rotorcraft       410    0.12     100.00
9         type                    TOTAL    336776      NA         NA
10      engine                  4 Cycle        48    0.01       0.01
11      engine                       NA     52606   15.62      15.63
12      engine            Reciprocating      1774    0.53      16.16
13      engine                Turbo-fan    240915   71.54      87.70
14      engine                Turbo-jet     40976   12.17      99.87
15      engine               Turbo-prop        47    0.01      99.88
16      engine              Turbo-shaft       410    0.12     100.00
17      engine                    TOTAL    336776      NA         NA
18 name_origin      John F Kennedy Intl    111279   33.04      33.04
19 name_origin               La Guardia    104662   31.08      64.12
20 name_origin      Newark Liberty Intl    120835   35.88     100.00
21 name_origin                    TOTAL    336776      NA         NA
22    dst_dest                        A    323811   96.15      96.15
23    dst_dest                        N      5363    1.59      97.74
24    dst_dest                       NA      7602    2.26     100.00
25    dst_dest                    TOTAL    336776      NA         NA
26  tzone_dest        America/Anchorage         8    0.00       0.00
27  tzone_dest          America/Chicago     74811   22.21      22.21
28  tzone_dest           America/Denver     10291    3.06      25.27
29  tzone_dest      America/Los_Angeles     46324   13.76      39.03
30  tzone_dest         America/New_York    192377   57.12      96.15
31  tzone_dest          America/Phoenix      4656    1.38      97.53
32  tzone_dest                       NA      7602    2.26      99.79
33  tzone_dest         Pacific/Honolulu       707    0.21     100.00
34  tzone_dest                    TOTAL    336776      NA         NA
35  lat_origin                40.639751    111279   33.04      33.04
36  lat_origin                  40.6925    120835   35.88      68.92
37  lat_origin                40.777245    104662   31.08     100.00
38  lat_origin                    TOTAL    336776      NA         NA
39  lon_origin               -73.778925    111279   33.04      33.04
40  lon_origin               -73.872608    104662   31.08      64.12
41  lon_origin               -74.168667    120835   35.88     100.00
42  lon_origin                    TOTAL    336776      NA         NA
43  alt_origin                       13    111279   33.04      33.04
44  alt_origin                       18    120835   35.88      68.92
45  alt_origin                       22    104662   31.08     100.00
46  alt_origin                    TOTAL    336776      NA         NA
\end{CodeOutput}
\end{CodeChunk}

We can also check for degree of association between the target variable which in this case is taken as "dst\_dest" i.e. day light savings time zone of destination with other variables such as origin, type, timezone and more. We can see in Table \ref{tab:sum3} that the degree of association between the target and the timezone variables is very high, which is expected.

\begin{CodeChunk}
\begin{CodeInput}
R> ExpCatStat(flight_data,Target="dst_dest",Label="Destination",
      result = "Stat",clim=10,nlim=5,Pclass="Yes")
\end{CodeInput}
\end{CodeChunk}

\begin{table}[!htp]
\centering
\caption{Summary of numerical variables of NYC flights data}
\label{tab:sum3}
\scalebox{0.65}{
\begin{tabular}{llllllllllllllllllll}
\toprule
Variable     & Target    & Unique & Chi-squared & p-value & df & IV Value & Cramers V & Degree of Association & Predictive Power \\ \midrule
origin       & dst\_dest & 3      & 2605.078    & 0       & 2  & 0        & 0.09      & Weak                  & Not Predictive   \\
type         & dst\_dest & 4      & 17.092      & 0       & 2  & 0        & 0.01      & Very Weak             & Not Predictive   \\
engine       & dst\_dest & 7      & 269.091     & 0       & 5  & 0        & 0.03      & Very Weak             & Not Predictive   \\
name\_origin & dst\_dest & 3      & 2605.078    & 0       & 2  & 0        & 0.09      & Weak                  & Not Predictive   \\
tzone\_dest  & dst\_dest & 8      & 329174      & 0       & 6  & 0        & 1         & Strong                & Not Predictive   \\
engines      & dst\_dest & 5      & 18.615      & 0       & 3  & 0        & 0.01      & Very Weak             & Not Predictive   \\
lat\_origin  & dst\_dest & 3      & 2605.078    & 0       & 2  & 0        & 0.09      & Weak                  & Not Predictive   \\
lon\_origin  & dst\_dest & 3      & 2605.078    & 0       & 2  & 0        & 0.09      & Weak                  & Not Predictive   \\
alt\_origin  & dst\_dest & 3      & 2605.078    & 0       & 2  & 0        & 0.09      & Weak                  & Not Predictive                \\ \bottomrule
\end{tabular}}
\end{table}

\begin{figure}[!htp]
    \centering
    \subfloat[Scatter plot of departure delay vs. arrival delay]{{\includegraphics[width=0.5\textwidth, keepaspectratio]{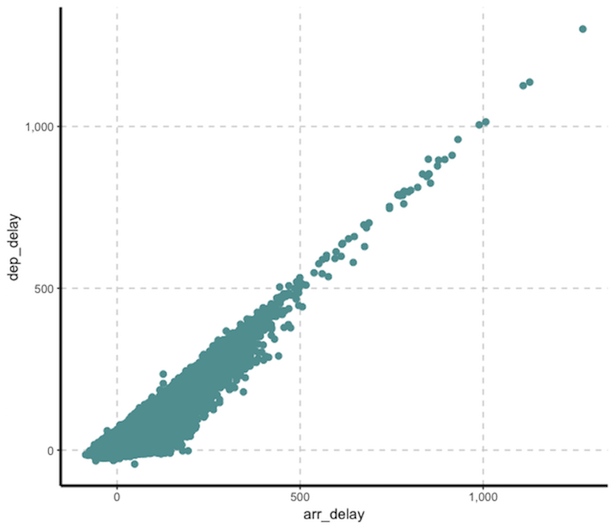} }}%
    \qquad
    \subfloat[Density plot for number of Seats]{{\includegraphics[width=0.5\textwidth, keepaspectratio]{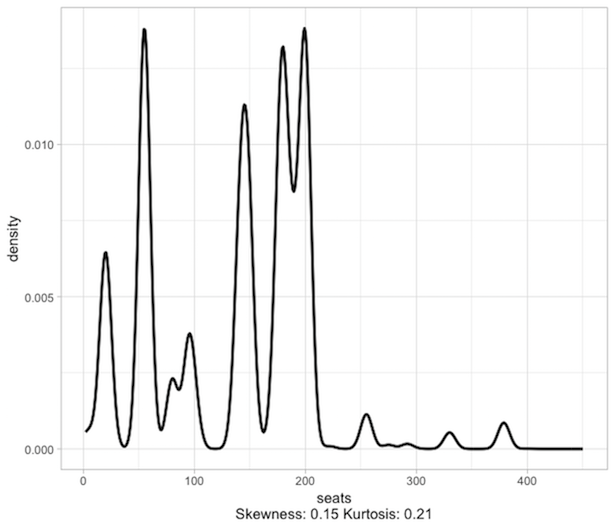} }}%
    \caption{EDA on NYC Flights data: Scatter plot and Density plot}%
    \label{fig:nyc1}%
\end{figure}

\begin{figure}[!htp]
    \centering
    \subfloat[Bar plot for number of Engines]{{\includegraphics[width=0.5\textwidth, keepaspectratio]{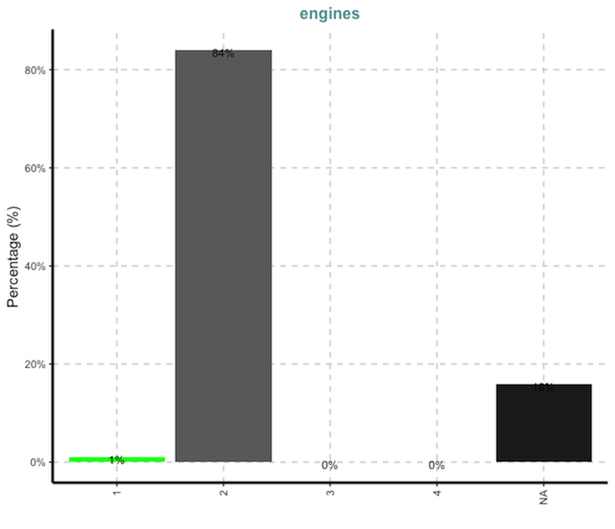} }}%
    \qquad
    \subfloat[Box plot of distance vs. origin]{{\includegraphics[width=0.5\textwidth, keepaspectratio]{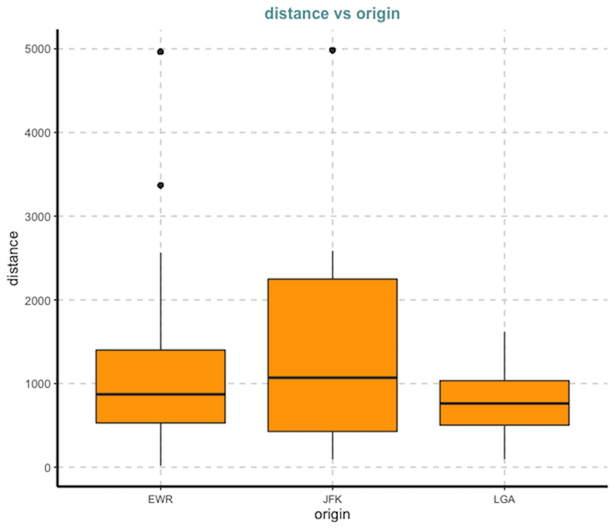} }}%
    \caption{EDA on NYC Flights data: Bar plot and Box plot}%
    \label{fig:nyc2}%
\end{figure}

\begin{figure}[!htp]
    \centering
    \subfloat[Normality plot for departure delay]{{\includegraphics[width=0.5\textwidth, keepaspectratio]{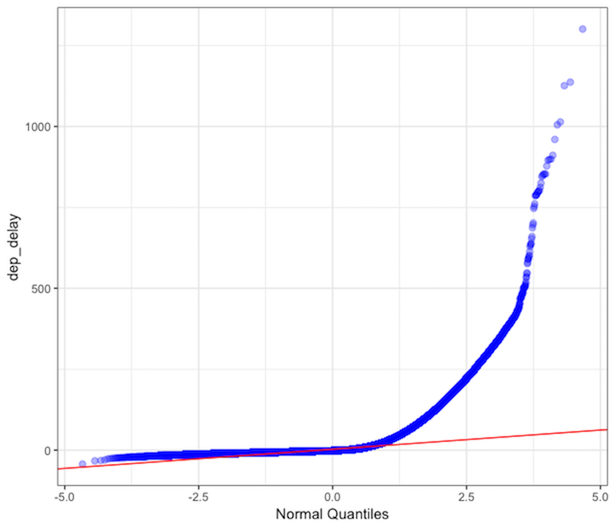} }}%
    \qquad
    \subfloat[Co-ordinate plot for departure delay, number of seats, arrival delay, air time and distance with respect to the origin of flights]{{\includegraphics[width=0.5\textwidth, keepaspectratio]{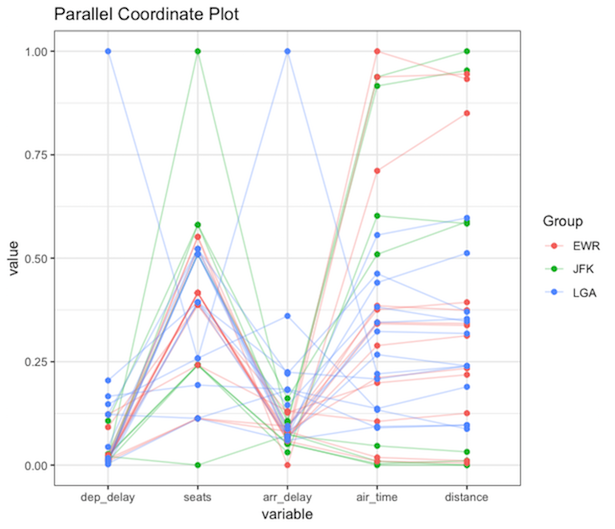} }}%
    \caption{EDA on NYC Flights data: Normality plot and Co-ordinate plot}%
    \label{fig:nyc3}%
\end{figure}

We can visualize the different graphical representations of the data using the SmartEDA package. Figures \ref{fig:nyc1}, \ref{fig:nyc2} and \ref{fig:nyc3} show the different graphical visualizations (such as Scatter plot, Density plot, Bar plot, Box plot, Normality plot and Co-ordinate plot) applied on the NYC flights dataset. Some of the key insights that we can get from the Figures \ref{fig:nyc1}, \ref{fig:nyc2} and \ref{fig:nyc3} are that there is a linear relationship between departure and arrival delays. Most of the flights have $2$ engines. Most of the longer distance flights are originating from the JFK airport. The variable "departure delay" is not normally distributed. 

\begin{CodeChunk}
\begin{CodeInput}
## Scatter plot
R> ExpNumViz(flight_data[,c("dep_delay","arr_delay")],gp="arr_delay",
      nlim=40,fname=NULL,col=NULL,Page=NULL,sample=1)
## Density plot
R> ExpNumViz(flight_data,gp=NULL,nlim=10,sample=1) 
## Bar plot
R> ExpCatViz(flight_data,gp=NULL,clim=5,margin=2,sample=1) 
## Box plot
R> ExpNumViz(flight_data,gp="origin",type=2,nlim=10,sample=1) 
## Normality plot
R> ExpOutQQ(flight_data,nlim=10)
## Co-ordinate plots
R> ExpParcoord(flight_data,Group="origin",Stsize=c(10,15,20),
      Nvar=c("dep_delay","seats","arr_delay","air_time","distance"))
\end{CodeInput}
\end{CodeChunk}

\section{Comparison with other packages} \label{compare}
Table \ref{tab:pkgcompare} compares the SmartEDA package \citep{smartEDA} with other similar packages available in CRAN for exploratory data analysis viz. dlookr \citep{pack:1}, DataExplorer \citep{pack:2}, Hmisc \citep{pack:3}, exploreR \citep{pack:4}, RtutoR \citep{pack:5} and summarytools \citep{pack:6}. The metric for evaluation is the availability of various desired features for performing an Exploratory data analysis such as (a) Describe basic information for input data, (b) Function to provide, (c) summary statistics for all numerical variables, (d) Function to provide plots for all numerical variables, (e) Function to provide summary statistics for all character or categorical variables, (f) Function to provide plots for all character or categorical variables, (g) Customized summary statistics- extension to data.table package, (h) Normality/ Co-ordinate plots, (i) Feature binarization/ binning, (j) Standardize/ missing imputation/ diagnose outliers and (k) HTML report using rmarkdown/ Shiny.


\begin{table}[!htp]
\centering
\caption{Comparison with available R packages}
\label{tab:pkgcompare}
\scalebox{0.62}{
\begin{tabular}{lccccccc}
\toprule
\textbf{Exploratory analysis features}                                                                                                & \textbf{SmartEDA} & \textbf{dlookr} & \textbf{DataExplorer} & \textbf{Hmisc} & \textbf{exploreR} & \textbf{RtutoR} & \textbf{summarytools} \\ \midrule
Describe basic information for input data                                                                                    & Y        & Y      & Y            & Y     &          &        & Y            \\\noalign{\smallskip}\hline
\begin{tabular}[c]{@{}l@{}}Function to provide summary statistics \\ for all numerical variable\end{tabular}                 & Y        & Y      &              & Y     & Y        &        & Y            \\\noalign{\smallskip}\hline
\begin{tabular}[c]{@{}l@{}}Function to provide plots for all numerical \\ variable\end{tabular}                              & Y        &        & Y            &       &          &        & Y            \\\noalign{\smallskip}\hline
\begin{tabular}[c]{@{}l@{}}Function to provide summary statistics \\ and plots for all character or categorical\end{tabular} & Y        & Y      & Y            & Y     &          &        &              \\\noalign{\smallskip}\hline
\begin{tabular}[c]{@{}l@{}}Function to provide plots for all \\ character or categorical\end{tabular}                        & Y        &        & Y            &       &          &        & Y            \\\noalign{\smallskip}\hline
\begin{tabular}[c]{@{}l@{}}Customized summary statistics - \\ extension of data.table package\end{tabular}                   & Y        &        &              &       &          &        &              \\\noalign{\smallskip}\hline
Normality  / Co-ordinate plots                                                                                               & Y        & Y      & Y            &       &          &        &              \\\noalign{\smallskip}\hline
Feature binarization / Binning                                                                                               &          & Y      & Y            &       &          &        &              \\\noalign{\smallskip}\hline
\begin{tabular}[c]{@{}l@{}}Standardize /missing imputation / diagnose \\ outliers\end{tabular}                               &          & Y      & Y            &       & Y        &        &              \\\noalign{\smallskip}\hline
HTML report using rmarkdown / Shiny                                                                                          & Y        & Y      & Y            &       &          & Y      &                         \\ \bottomrule
\end{tabular}}
\end{table}

We can see in Table \ref{tab:pkgcompare} that the current version of SmartEDA has almost all the desired characteristics mentioned above except the points (h) and (i) i.e. normality plots and feature binning respectively. These two features would be incorporated in the next release and we are currently working on it. However, the unique and the strongest functionality provided by SmartEDA is point (f) i.e. extension to data.table package which none of the other packages provide. Thus, SmartEDA does add value given the importance and popularity of data.table among R users for analyzing large datasets. Table \ref{tab:pkgcompare} shows that SmartEDA is better than almost all the other packages available in CRAN. The closest competitor to SmartEDA seems to be the DataExplorer package but it doesn't possess the (b) and (f) features viz. Function to provide summary statistics for all numerical variables and extension to data.table package respectively.

\section{Conclusion} \label{con}
The contribution of this paper is in development of a new package in R i.e. SmartEDA for automated Exploratory Data Analysis. SmartEDA package helps in implementing the complete Exploratory Data Analysis just by running the function instead of writing lengthy R code. The users of SmartEDA can automate the entire EDA process on any dataset with easy to implements functions and export EDA reports that follows the industry and academia best practices. The SmartEDA can provide summary statistics along with graphical plots for both numerical and categorical variables. It also provides extension to data.table package which none of the other packages available in CRAN provides. Overall, the main benefits of SmartEDA are in development time savings, less error percentage and reproducibility.  As of March 2019, the SmartEDA package has more than 3100+ downloads, which indicates its acceptability and maturity in the Statistics and Machine learning community.

Some of the current limitations of the SmartEDA are that it cannot perform variable transformation, feature engineering and dynamic visualization. Another area for further development includes functions to generate automated shiny dashboard with some standard templates for some basic EDA presentation. We are working on adding these functionalities in the future releases of SmartEDA.


\section*{Acknowledgments}
We would like to thank VMware and the Enterprise \& Data Analytics (EDA) leadership for giving us the required infrastructure and support for this work. We are grateful to the R community for their acceptance and feedback to improve our package further.


\bibliography{refs}


\newpage


\end{document}